# Overview on the physics and materials of the new superconductor $K_xFe_{2-y}Se_2$

Hai-Hu Wen

Center for Superconducting Physics and Materials, National Laboratory for Solid State Microstructures, Department of Physics, Nanjing University, Nanjing 210093, China

E-mail: hhwen@nju.edu.cn

Since the discovery of high temperature superconductivity in iron pnictides in early 2008, many iron-based superconductors with different structures have been discovered, with the highest transition temperature to date being 57 K. By the end of 2010, another kind of new superconductor, the Fe-based chalcogenide $K_{1-x}Fe_{2-y}Se_2$ was discovered. A naive counting of the electrons in the system would lead to a conclusion that the system is heavily electron overdoped (~0.4 e/Fe). Band structure calculations further support this speculation and predict that the hole pockets which are found in the iron pnictides may be missing. This greatly challenges the widely perceived picture that the superconducting pairing is established by exchanging anti-ferromagnetic spin fluctuations and the electrons are scattered between the electron and hole pockets. Later, it was found that both potassium and iron might be deficient in $K_{1-x}Fe_{2-y}Se_2$, yielding to a picture of phase separation. In this picture the superconducting phase and the antiferromagnetic (AF) phase may phase separate spatially into different regions. This generates further curiosity about what is the real superconducting phase, what is the relationship between the superconducting phase and the AF phase, and what is the parent state for the superconducting phase. We propose a "spider web" model for the phase separation, which can explain both the transport and magnetic data. In this paper, we review the status of research in this rapidly growing field and list the important and unsettled issues as perspectives for future research.

# 1. Introduction

Superconductivity is achieved through the quantum condensation of Cooper pairs that are charge carriers formed by two bound electrons. To form a paired electron state with a strong binding force is the core issue to have high temperature superconductivity. The discovery of high temperature superconductivity in LaFeAsO$_{1-x}$F$_x$ greatly stimulated efforts towards this direction.[1] The consensus in the community is that the family of iron pnictides is another high temperature superconducting system, after the cuprates, and the electron pairing is not established through conventional electron-phonon coupling. In the three years since the discovery of LaFeAsO$_{1-x}$F$_x$, several different structures have been discovered, called now in short as 11, 111, 122, 1111, 32522 and 21311 (or 42622), as described in several review papers describing the materials and properties of the iron pnictides. [2, 3, 4, 5, 6, 7, 8] It is clear that all these families have FeAs or FeSe planes as the common building blocks, which are responsible for the superconductivity. Theoretically it is suggested that the pairing interaction is mediated by exchange of the antiferromagnetic (AF) spin fluctuations, and the pairing is due to the hopping of electrons between the electron and hole pockets. [9, 10] A natural consequence of this pairing manner is the so-called S$^\pm$ order parameter symmetry, where full gaps (with a certain anisotropy) appear on the electron and hole pockets but with opposite signs. Intuitively, this picture requires a reasonably good nesting (however, should not be perfect, otherwise the spin-density wave phase may prevail) between the electron and hole pockets. The inelastic neutron scattering finds a resonance peak of the imaginary part of the electron spin susceptibility at the momentum Q = [$\pi$, $\pi$] (in the folded Brillouin Zone notation) which measures exactly the AF wave vector as well as the Fermi surface nesting vector (connecting the hole-electron pockets) in the FeAs-based 122 and 1111 system [11,12]. However, in the FeSe-based 11 system the in-plane magnetic moment aligns 45 degrees from the Fe-Fe bond direction. In this case the AF wave vector is (0, $\pi$), which does not match the inter-electron-hole pocket wave vector ($\pi$, $\pi$). Even in this case, inelastic neutron scattering reveals a resonance peak that is consistent with the inter-electron-hole pocket wave vector, supporting the picture of $S^{\pm}$ order parameter symmetry.[13]

By the end of 2010, one group in the Institute of Physics at the Chinese Academy of Sciences reported an interesting discovery: they found superconductivity at about 32 K in $K_xFe_2Se_2$.[14] Almost simultaneously, superconductivity was also found in $Rb_xFe_2Se_2$ [15] and $Cs_xFe_2Se_2$ [16]. The temperature dependence of the resistivity and the magnetic susceptibility are shown in Fig.1 for $Rb_{0.8}Fe_{1.6}Se_2$. According to the early paper by Guo et al. [14], the structure of this new superconductor is the same as $ThCr_2Si_2$, the so-called 122 structure. However, there are several distinctions if we compare this new superconductor with the FeAs-based 122 (e.g., $Ba_{1-x}K_xFe_2As_2$ or $Ba(Fe_{1-x}Co_x)_2As_2$): (1) The normal state resistivity exhibits a huge hump in the temperature region of 100-300 K. This hump can move to a different temperature depending on post-annealing or applied pressure; (2) The low temperature resistivity is much larger and is sensitively dependent on growth and heat treating conditions; (3) The superconducting volume, as judged from magnetic shielding at high applied field, is only about 20%. As mentioned earlier, a simple counting of the electrons based on the valence considerations indicates that the sample should be electron overdoped. As potassium is highly reactive it is naturally assumed that the potassium is deficient, which is supported by the energy-dispersive spectrum analysis results. A band structure calculation indicates that the system is indeed heavily electron overdoped and the hole pocket may sink below the Fermi energy [17]. In this case, the prerequisite for the pairing scattering in the $S^{\pm}$ model is not satisfied since the hole pocket is missing.

## 2. Fe-vacancy orders and the block antiferromagnetic state

The original FeSe-based 122 phase, $Tl_xFe_{2-y}Se_2$ was discovered in 1978, and neutron scattering data revealed Fe vacancy structures.[18] This message was brought to us by Fang et al.[19]. They also found superconductivity in $(Tl,K)Fe_2Se_2$, and speculated two kinds of Fe-vacancy orders, as shown in Fig.2. In one structure with the formula of $K_xFe_{1.5}Se_2$, the iron atoms have two or three iron neighbors. In the other structure with the formula of $K_xFe_{1.6}Se_2$ (now called the $A_2Fe_4As_5$ phase), each iron atom has three iron neighbors, which was later shown to be a very commonly formed structure but gives a unique AF structure. In both of these phases, Fang et al [19] showed that the potassium concentration can be simply calculated by charge balance analysis when assuming that the Fe has a "2+" ionic state.. This group also pointed out that the "parent phase" is an insulator, which was confirmed experimentally by several groups preparing the nominal composition of $KFe_2Se_2$ . This parent phase has not been found to exhibit superconductivity. This insulating phase was shown to be the Fe-vacancy ordered 245 phase. With the two kinds of blocks of Fe-vacancies suggested by

Fang et al.[19], band structure calculations indeed indicated a band gap with the value of about 0.3-0.5 eV for $K_{1-x}Fe_{1.5}Se_2$[20] and 60 meV for $K_{1-x}Fe_{1.6}Se_2$.[21] It would be very interesting to know whether the superconductivity is directly evolved from this insulating phase by filling up the Fe-vacanciesm or arises from disorder, as suggested from our quenching experiment[22].

In order to check which kind of the Fe vacancy order is most likely formed, Bao et al. performed neutron diffraction experiment[23]. Their structural refinement revealed that the dominating phase had a composition of $K_{0.82}Fe_{1.62}Se_2$, that corresponds to the second Fe-vacancy ordered structure reported by Fang et al.[19], namely the 245 phase. Furthermore Bao et al. also found a "block AF" state with 3.31 $\mu_B$/Fe. This magnetic structure is shown in Fig.3, where one can see that four nearby Fe ions form a square-like block with the spins aligned ferromagnetically. The total magnetic moment of the 4-iron-block is 13.24 $\mu_B$. This magnetic structure was later shown by the ORNL group that this is a common crystalline and magnetic structure of the "superconducting $A_2Fe_4Se_5$" (A=K, Rb, Cs and Tl) phase [24]. Bulk measurements using transport,magnetization and specific heat on these materials also showed a magnetic and structural transition at very high temperatures (470-560 K), corresponding to the formation of the block AF ordering.[23, 24, 25] This kind of Fe-vacancy order was also observed by transmission electron microscopy[26].

## 3. Phase separation and hunting for the the superconductiviting phase

Since the Fe vacancy and its order are commonly observed, and the band structure calculation clearly indicates that there is a band gap at the Fermi energy, it is puzzling why this phase gives rise to superconductivity. On the other hand, many experiments indicate the existence of an antiferromagnetic order in materials exhibiting superconductivity. Typical Mössbauer data clearly indicate the formation of an AF state below 556 K.[27] Electron backscattering image analysis of a cleaved surface finds mesoscopic inhomogeneity, which shows rather uniform areas enclosed by a network of thin random strips with width less than 1 μm. Detailed analysis show that the paramagnetic phase constitutes only about 12 % of the surface area. As mentioned above, the block AF structure would lead to a magnetic moment of 13.24 $\mu_B$. As argued by Mazin[28], Cooper pairs with singlet pairing will be broken in such a strong local magnetic field, even if with a canting angle of only 0.05 degrees.

How can we reconcile these phenomena with such contradictions? One way is to assume the picture of phase separation. The system, due to either chemical or electronic causes, separates into the block AF and the superconducting phases. Due to the volatility of the potassium, chemical inhomogeneity is likely, but it remains an open issue whether this potassium inhomogeneity influences the electronic uniformity in the neighboring layers. This phase separation picture was quickly supported by the magnetization measurements[29]. In Fig.4, we present the comparison of the magnetization-hysterisis-loop curves of the $K_{0.8}Fe_{1.6}Se_2$ single crystals and the $Ba_{0.6}K_{0.4}Fe_2As_2$ and $Ba(Fe_{0.92}Co_{0.08})_2As_2$ (FeAs-122). There are three clear distinctions between them. First, the full magnetic penetration (when the magnetic flux fronts meet at the center of the sample) to the Meissner state in $K_{0.8}Fe_{1.6}Se_2$ occurs at a much lower field than in FeAs-122. The full magnetic penetration field is only about 300 Oe, while that of FeAs-122 is at least 3000 Oe, indicating magnetic flux penetrates into the $K_{0.8}Fe_{1.6}Se_2$ much more easily. Second, the width of the magnetization hysteresis loops (which is proportional to the critical current density in the Bean critical state model) is about 100 times smaller in $K_{0.8}Fe_{1.6}Se_2$ than in $Ba_{0.6}K_{0.4}Fe_2As_2$, and 50 times smaller than in $Ba(Fe_{0.92}Co_{0.08})_2As_2$, indicating that the critical current density is much smaller in $K_{0.8}Fe_{1.6}Se_2$. Third, the magnetization hysteresis loops exhibit a valley, instead of a peak, near zero field, clearly indicating a low critical current density near zero field. The magnetic shielding at a low field, as measured by other groups in the zero-field-cooled M(T) curves, can reach 100%, and it has been claimed that the superconductivity volume is 100 %. We point out that in cases of non-uniform morphology, shielding will not provide the superconducting volume fraction – and a true Meissner (for example, the field cooled magnetization ) experiment must be performed. The fact that the shielding percentage drops dramatically under an increased magnetic field (> 0.1 T), indicates this non-uniformity, supporting the phase separation model. In exploring this, we performed a series of experiments where the as-grown single crystals were quenched, to see if superconductivity could be induced. After a thermal quenching at about 500-600 K, the crystals did exhibit superconductivity at low temperatures.[22] This experiment suggests two possibilities: (1) The superconducting phase has the 245 structure, but the Fe vacancies are in a disordered state. From the simple band structure point of view, disorder will induce some density of states in the band gap, which may lead to superconductivity. Based on the Hubbard model the Rice University group [30] calculated the phase diagram of the 245 system with Fe vacancy order and disorder. They found that the phase with disordered Fe-vacancies had a lower free energy and a finite density of states at the Fermi energy; (2) The morphology of the crystals is such that non-superconducting 245 phase is complexly enclosed by the superconducting phase (in this case could be Fe vacancy free or less). This structure resembles that of a "spider web," where the strings of the spider web take up only very small volume, but are very strong. This model is further supported by the robust superconductivity revealed by the resistivity in high magnetic fields.

Other measurements support the phase separation model,. But direct evidence comes from scanning tunneling microscopy. Li et al. grew [110]-oriented $K_xFe_{2-y}Se_2$ thin films and found the superconducting phase to be Fe vacancy-free with the standard formula $KFe_2Se_2$[31], surrounded by the state with the block AF state ($\sqrt{5}\times\sqrt{5}$). This supports the second possibility as outlined above, with the superconducting phase being heavily electron overdoped. The STM data this directly show that the films contain two different regions which are microscopically phase separated. This however is inconsistent with the robust superconductivity as measured by the resistive transitions under a magnetic field. As shown in Fig.5, the general feature of the resistive transitions under a high magnetic field is quite similar to that of $Ba_{1-x}K_xFe_2As_2$ or $Ba(Fe_{1-x}Co_x)_2As_2$ : The field induced broadening of the transition is not as large as that in a superconductor with spatial percolation or diluted superfluid density. Consider that the phase separation is microscopic in origin, consisting of many tiny Josephson weak links. It can be theoretically treated as a 2D XY-model, and the system should exhibit strong superconducting phase fluctuations. This could cause electronic phase separation (as opposed to the chemical or crystallographic phase separation discussed above) also with a "spider web" structure. This "spider web" model allows significant magnetic penetration, while the superconducting paths or filaments can carry the supercurrent under even a very high magnetic field.[32] Since the volume fraction of the superconducting filaments is only 15-20 %, mapping out any phase diagrams (versus doping) from the bulk or even local TEM measurements, will suffer from big uncertainties.

**4. Pairing symmetry and the gap structure**

Due to the nature of phase separation, it is however still too early to declare any solid evidence for the symmetry of the superconducting order parameter. The angle resolved photo-emission spectroscopy (ARPES) data from three different groups all reveal an isotropic gap structure, with the hole pocket appearing in the FeAs counterpart, completely missing.[33,34,35] The ARPES data do indicate a mixture of two different phases: one with a finite DOS at the Fermi energy that may be the one responsible for superconductivity; and the other with some bands that are about 0.5 eV away from the Fermi energy (note that the distance of the top of the hole band at $\Gamma$ is only 75 meV from the Fermi energy), and this valence band could arise from the insulating phase.[36] Other tools for probing the pairing symmetry, like the thermal conductivity and the penetration depth measurements, are still lacking in $K_xFe_{2-y}Se_2$. This is due to the difficulty in obtaining meaningful data in the presence of phase separation effects. Both the thermal conductivity and penetration depth measurements detect not only the quasi-particle (QP) density of states at the Fermi energy, but also the "dynamics" of these QPs, namely the scattering rate and the Fermi velocity, and it is not known how to analyze the data in the presence of if two phases. Specific heat measurements, however, detect only the magnitude of the DOS at the Fermi energy. The specific heat data measured at low temperatures indicate that the superconducting phase is quite pure as shown by a quite sharp anomaly at $T_c$. The specific heat anomaly near $T_c$ is about 10 mJ/molK$^2$[37]. This is slightly lower than that in the $SmFeAsO_{1-x}Fe_x$ system with $T_c$

= 54 K [38], but is only 1/10 of the value in $Ba_{0.6}K_{0.4}Fe_2As_2$ [38]. If we assume that the superconducting volume is only about 20%, the specific heat anomaly may be as large as 50 mJ/molK$^2$ for the pure superconducting phase, this high value suggests a strong electron-boson coupling. The field dependence of the low temperature specific heat normally tells how many Cooper pairs are broken by the magnetic field and the gap structure. For an isotropic gap, the field-induced enhancement of the specific heat coefficient should be linear with field. Preliminary experiments do reveal a linear field dependence suggesting a full, isotropic gap. Interestingly, STM measurements [31] on the superconducting area at 0.4 K find two gaps: $\Delta_1$ = 4 meV, $\Delta_2$ = 1 meV, and there is a clear "V" shape in the tunneling conductance near zero energy. Whether this suggests a nodal gap structure is still an unresolved issue. It would be very interesting to measure the tunneling spectrum in the superconducting area of a naturally cleaved single crystal, and to resolve the gap structure by using the quasiparticle interference pattern.

Recently, inelastic neutron scattering experiments in the system of $Rb_{1-x}Fe_{2-y}Se_2$ demonstrate a magnetic resonance peak at the (0.5,0.25,0.5) with a resonance energy of about 14 meV.[39,40] This neutron resonance is explained within the AF spin-fluctuation mediated inter-pocket scattering picture, but the pairing scattering occurs between the two nearby electron pockets. Because the real part of the non-interacting spin susceptibility is strongest when the Fermi velocity at the initial and final states of the scattering is opposite in directions, considering the slightly square-like electron Fermi pockets, the resonance occurs not at ($\pi,\pi$) but at ($\pi,\pi$)±$k_F$, where $k_F$ is the momentum of the Fermi surface measured from the center of the electron pocket. Figure 6 shows the data and a sketch of this picture based on these experiments. This experiment, if interpreted correctly, strongly suggests that there should be a sign change of the superconducting gap either between the neighboring electron pockets or within one pocket. Although this result is not supported by measurements using other experiment tools so far, for example the ARPES data, it certainly warrants further investigation. Recently, Hirschfeld, Korshunov, and Mazin gave a nice summary of the present status in determining the gap structure based on theoretical and the structural considerations. [8] They conclude that the quasi-nodeless d-wave, the incipient S±, and the bonding-antibonding S±, are the three most possible candidates. The first and the third should exhibit interesting gap sign reversal.

## 5. Conclusions and perspectives

Apparently the story about $K_xFe_{2-y}Se_2$ does not come to the end. There are at least several key issues left:

(1) What is the true chemical phase for the superconductivity? If it has the stoichiometric formula $KFe_2Se_2$, can it exist alone or does it rely on the neighboring non-superconducting phase, like the 245 phase, in order to overcome the charge balance problem in the nanoscale?

(2) What is the relationship between the superconducting phase and the block-AF phase with the $\sqrt{5}\times\sqrt{5}$ structure. Do they coexist in a microscopic scale or separate spatially? So far many experiments support the picture of phase separation.

(3) What is the parent phase of the superconductivity? Or we pose the question in a different

way: Does the superconductivity originate from some antiferromagnetic phase, or just like that in FeSe, without a parent AF state? Recently an interesting Fe-vacancy free block-AF state was proposed for the parent phase,[41] but this is still under debate on how the Cooper pairs can survive if a huge local magnetic moment exists assuming a slight canting of the magnetic moment of each Fe ion on these ferromagnetic blocks.

(4) In the very first paper about the superconductivity in $KFe_2Se_2$, a superconducting phase with $T_c$ of about 43 K was claimed and was later observed by other groups[14]. Recently, using a high pressure, superconductivity at about 48 K was reported [42], although the starting material had a bulk $T_c$ of 32 K with the nominal formula as $K_{0.7}Fe_{1.6}Se_2$. It is thus very curious to know the real phase that exhibits superconductivity above 40 K. More recently, several groups tried to insert the neutral charge component K-ammonide or Ba-ammonide between the FeSe layers and found superconductivity above 40 K[43,44]. These experiments may suggest that the real superconducting phase is Fe-vacancy free. This may explain why the single, perfect FeSe layer on the $SrTiO_3$ substrate exhibits superconductivity at such a high temperature ($T_c^{onset} = 53K$) [45]. In this work, the authors measure a pair of peaks which resemble that of the superconducting coherence peaks in a tunneling conductance spectrum. If these peaks really arise from the superconducting coherence, then superconductivity at much higher temperatures, such as 77 K (as claimed by the authors), is possible.

(5) Upon the determination of the chemical formula and the structure of the superconducting phase, the important issue of the pairing symmetry should be addressed. Great curiosity exists about whether the initial idea of the pairing manner via the inter-hole-electron-pocket scattering (S± model) survives or not. Even if the hole pockets are really missing, the S± picture seems to still be valid in describing the origin of superconductivity in the FeAs-baesd systems. The superconductivity in the FeAs-based and the FeSe-based systems may be generalized by the picture of AF spin fluctuation mediated pairing.

**Acknowledgements**

This work was performed in the support of NSF of China (11034011/A0402), the Ministry of Science and Technology of China (973 projects: 2011CBA00102, 2012CB821403). We thank Minghu Fang, Wei Bao and D. S. Inosov for allowing us to use part of the figures in their published papers. We also appreciate the useful discussions with Pengcheng Dai and Jiangping Hu.

**Figures and legends**

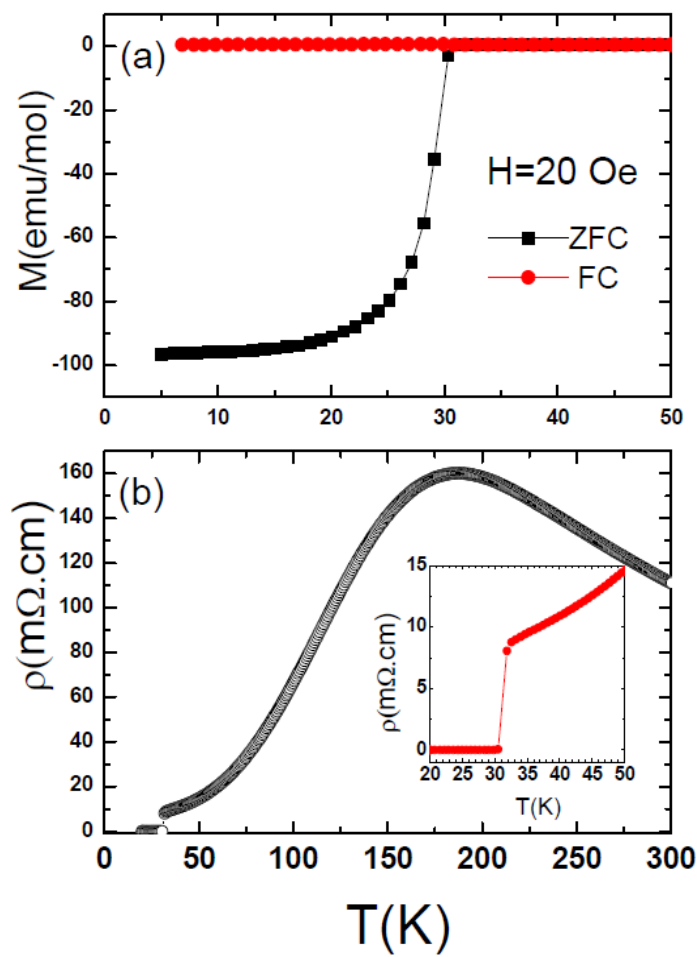

**Figure 1** (a) Temperature dependence of the DC magnetization measured in the zero-field-cooled and the field-cooled processes. One can see that the magnetic screening is very large (almost 100% after counting the demagnetization factor) at 20 Oe. This case will change dramatically when the external field is beyond about 0.1 T. (b) Temperature dependence of resistivity at zero field. The insert shows an enlarged view of the same data.

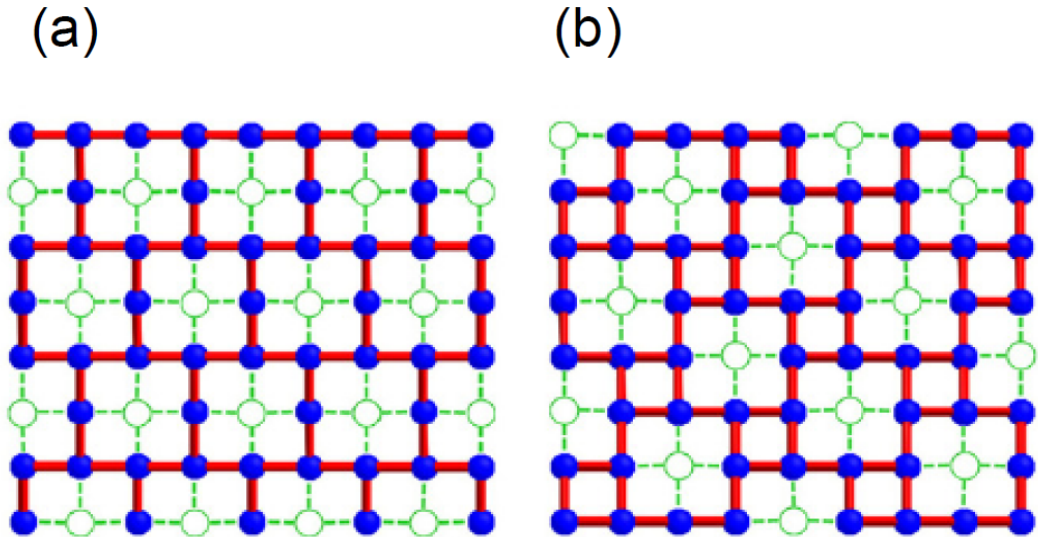

**Figure 2** Two possible Fe-vacancy ordered state corresponding to (a) $K_xFe_{1.5}Se_2$ (or called as the 234 phase) and (b) $K_xFe_{1.6}Se_2$ (or called as 245 phase if x takes 0.8). This is adopted from Ref.[19]. The solid blue circles represent the Fe ions. The open green circles stand for Fe-vacancies.

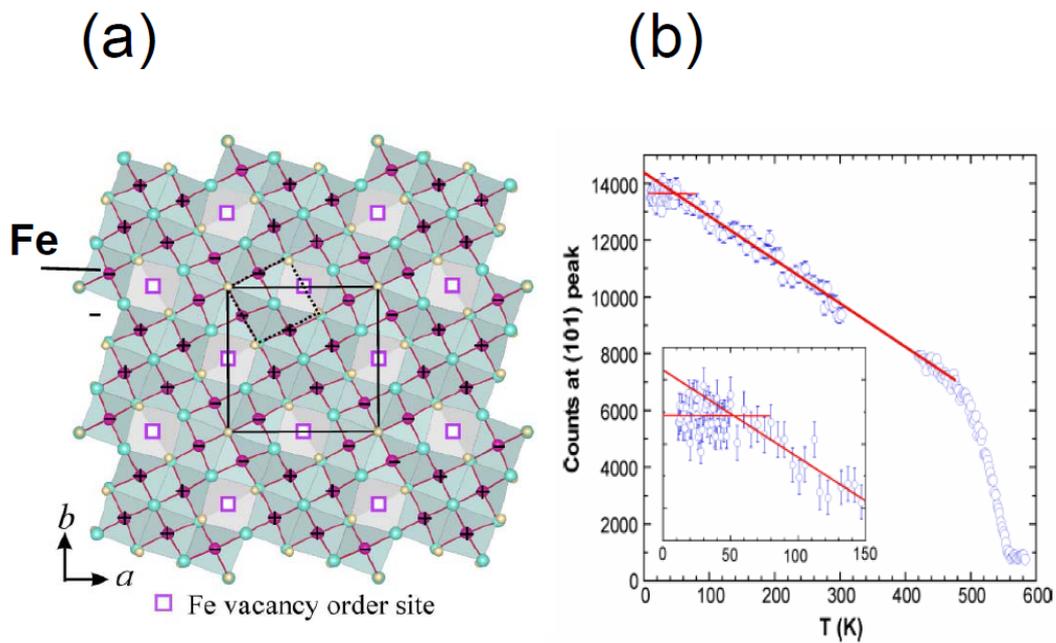

**Figure 3** (a) The so-called $\sqrt{5}\times\sqrt{5}$ block AF phase. One can see that the magnetic moments of four neighboring iron ions (purple circles) align them ferromagnetically, forming a unit with huge magnetic moment (13.4$\mu_B$). The pink open squares here represent the Fe-vacancies. (b) The squared magnetic order parameter. The magnetic Bragg peak (101) appears below the Neel temperature $T_N$ = 559 K. Below 500K, the Bragg intensity decreases linearly with temperature. The inset enlarges the low temperature part around $T_c$. This figure is reconstructed based on the data in Ref.[23]

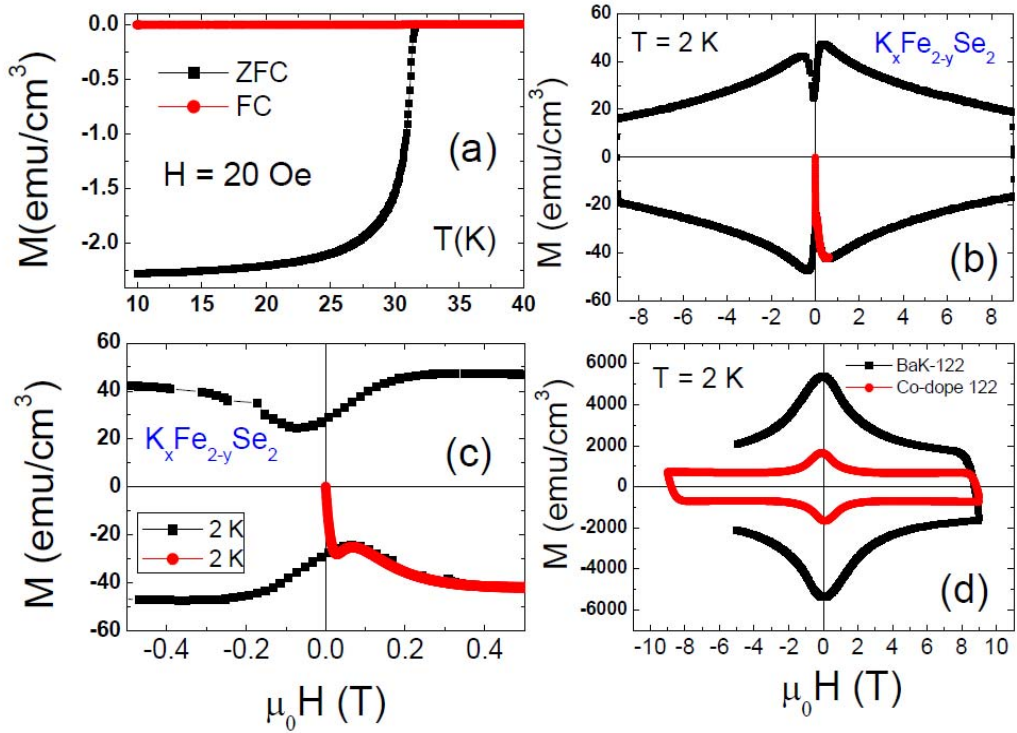

**Figure 4** (a) The temperature dependence of magnetization measured at H =20Oe with the field-cooling and zero-field-cooling processes. (b) The MHL measured at T = 2K and (c) an enlarged view in the low-field region during the magnetic-field penetration. The dark squares represent the data measured from 9T to −9T and back to 9T at a field sweeping rate of 200Oe/s. The red circles show the data in the low-field region measured with a rate of 5 Oe/s. An abnormal dip of MHL appears near zero field. (d) The MHLs measured at T = 2K for $Ba_{1-x}K_xFe_2As_2$ and $Ba(Fe_{1-x}Co_x)_2As_2$. Clearly a maximum appears near zero field for both samples, in contrast with that in the samples $K_xFe_{2-y}Se_2$. This figure is adopted from the Ref.[29]

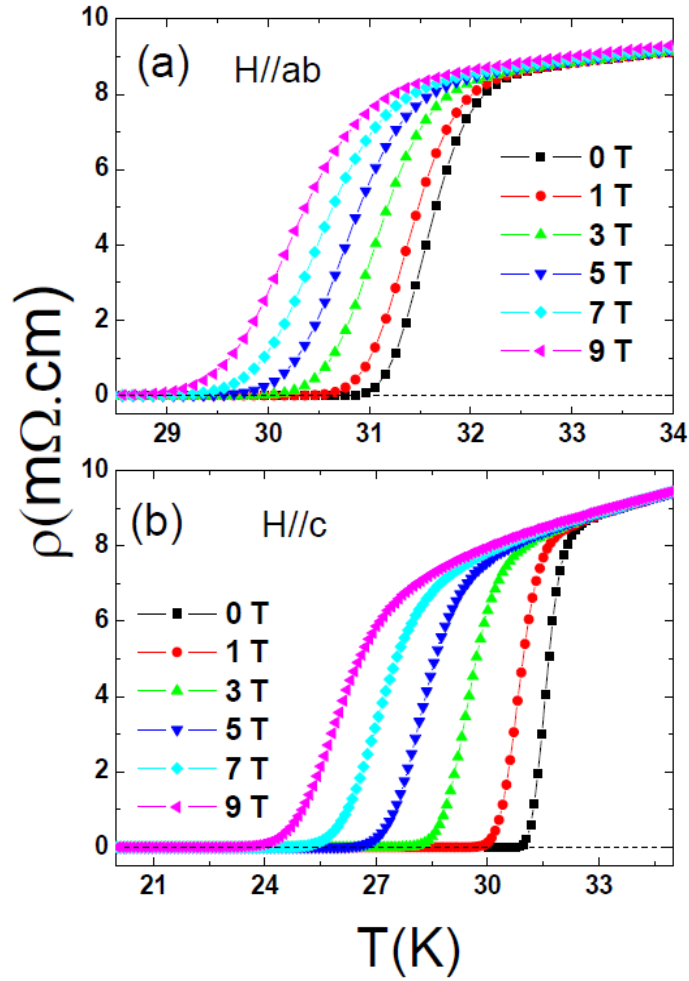

**Figure 5** The resistive transitions in Rb0.8Fe1.8Se2 with the magnetic field (a) parallel to the FeSe-planes and (b) perpendicular to the FeSe-planes. The magnetic field induced broadening of the transition is not large, actually very similar to that in $Ba_{0.6}K_{0.4}Fe_2As_2$ and $Ba(Fe_{0.092}Co_{0.08})_2As_2$. This suggests to us the strong filament superconductivity, which urges us to propose the "spider web" model.

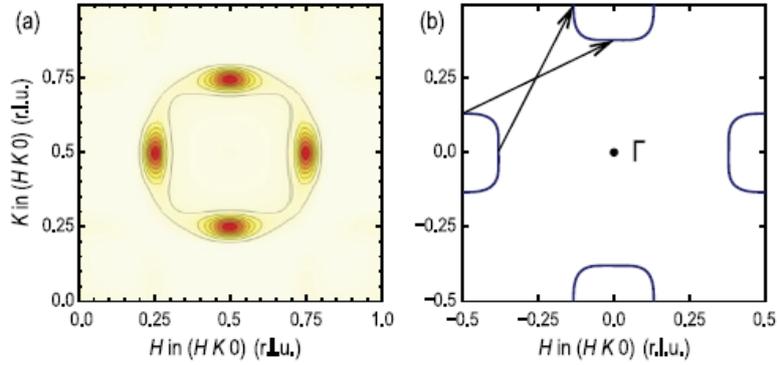

**Figure 6** (a) The neutron resonance spots in the q-space quoted from Ref.[39,40] for $Rb_{1-x}Fe_{2-y}Se_2$. The scattering spots locate around but not exactly at the $Q = (\pi, \pi)$. (a) Cartoon picture about the pairing scattering channels and the Fermi surfaces in the Brilloin zone, the scattering between the moment point with the opposite Fermi velocity is strongly favored.